\documentclass[aps,prl,twocolumn,groupedaddress,showpacs]{revtex4-1}

\usepackage{color}
\usepackage{amsmath}
\usepackage{graphicx}
\usepackage{bm}
\usepackage[normalem]{ulem}

\renewcommand{\eqref}[1]{Eq.~(\ref{#1})}

\begin{document}

\title{Membrane Viscosity Determined from Shear-Driven Flow in Giant Vesicles}

\author{Aurelia R. Honerkamp-Smith, Francis G. Woodhouse, Vasily Kantsler, and
 Raymond E. Goldstein}
\affiliation{Department of Applied Mathematics and Theoretical Physics, Centre for 
Mathematical Sciences,\\ University of Cambridge, Wilberforce Road, Cambridge CB3 0WA, 
United Kingdom}

\date{\today}

\begin{abstract}
The viscosity of lipid bilayer membranes plays an important role in determining the 
diffusion constant of embedded proteins and the dynamics of membrane deformations,
yet it has historically proven very difficult to measure.  
Here we introduce a new method based 
on quantification of the large-scale circulation patterns induced inside vesicles adhered to a solid surface 
and subjected to simple shear flow in a microfluidic device.  Particle Image Velocimetry based on 
spinning disk confocal imaging of tracer particles inside and outside of the vesicle, and 
tracking of phase-separated membrane domains are used to reconstruct the
full three-dimensional flow pattern induced by the shear.  These measurements 
show excellent agreement with the predictions of a recent theoretical analysis, and allow 
direct determination of the membrane viscosity.
\end{abstract}


\maketitle

Ever since the work of Saffman and Delbr{\"u}ck on the dynamics of inclusions in biological 
membranes \cite{SaffmanDelbruck} it has been recognized that lipid bilayers can be viewed as 
ultra-thin fluid layers endowed with a surface viscosity.  Along with that of the surrounding
fluid, this viscosity plays an important role in determining the translational and rotational 
diffusion constants of inclusions within the membrane \cite{HPW}.  A body of 
theoretical work \cite{KellerSkalak,NoguchiGompper} suggests that 
that nonequilibrium dynamics of 
vesicles in external flows  \cite{Kantsler} can also be sensitive to the value of this 
viscosity \cite{Seifert1993}. As the membrane viscosity $\eta_m$ can be expressed as 
$d\times \eta$, where $d$ is the
membrane thickness and $\eta$ is the bilayer fluid viscosity, the nanometric scale
of $d$ renders $\eta_m$ very small.  
Not surprisingly, it
has proven difficult to measure $\eta_m$; 
ingenious techniques that have been developed include measurements of 
the motion of membrane-bound microspheres \cite{Dimova},  
diffusion constants of domains in multicomponent membranes \cite{Petrov,Cicuta2007}, and observation of 
fluctuation dynamics in membranes near a critical point \cite{Camley,HonerkampSmith}.  Further afield, monolayers admit 
additional experimental techniques, including methods based on surface rheology \cite{Fuller2012} and microrheology
methods such as observing dynamics of submerged optically trapped \cite{Levine} or membrane-bound 
\cite{Sickert2007} microspheres.  Rheological experiments have the advantage of being able to detect 
non-Newtonian behavior \cite{Sadoughi2013}.

Interest in membrane dynamics also extends to flows {\it within} vesicles, especially in 
plant science, as the plant vacuole is 
contained within the vacuolar membrane (or {\it tonoplast}), which can comprise 
some of the largest lipid vesicles known: in 
internodal cells of the aquatic plant {\it Chara corallina} these can be cylinders $1$ mm in diameter
and up to $10$ cm long \cite{Protoplasma}.  This tonoplast is subject to 
continuous hydrodynamic shear through the action of cytoplasmic streaming, motion of 
the cytoplasm surrounding the vacuole \cite{Shimmen04}.
Because of its potential role in transport \cite{Goldstein2008}
there is great interest in the
three-dimensional characteristics of such shear-induced flows \cite{chara_jfm} and the role played by the
intervening tonoplast \cite{Wolff}.  

\begin{figure}[b]
\includegraphics[width = 0.95 \columnwidth]{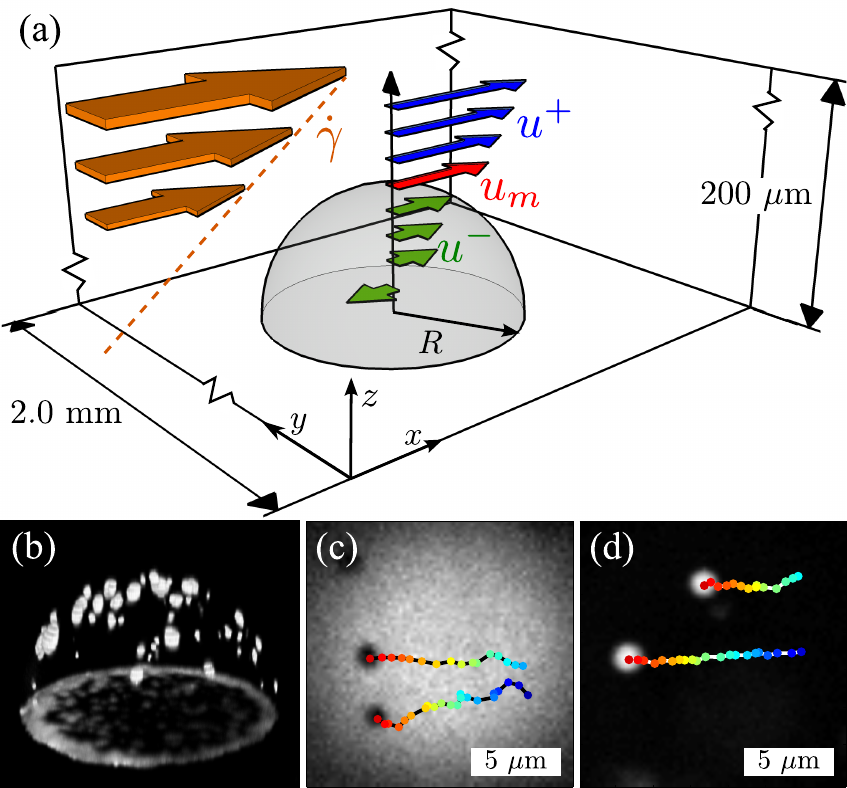}
\caption{(color online).  Microfluidic shear experiment.  (a) Schematic of the chamber 
(not to scale) and flows. (b) Confocal imaging reconstruction of an adhering 
hemispherical $L_{o}$ phase vesicle with small $L_{d}$ domains visible on its surface.  (c,d) Tracking of gel domains
in $L_{d}$ background (c) and $L_{d}$ domains in $L_{o}$ background (d), flowing across the vesicle 
apex at $\dot\gamma=2.6$ s$^{-1}$ (tracks color-coded in time over $\sim\! 2.6$ s).  }
\label{fig:droplet_diagram}
\end{figure}

A key development in the study of membrane fluid dynamics was the conceptually simple 
experiment of
V{\'e}zy, {\it et al.} \cite{Vezy} (see also \cite{Lorz}) in which a vesicle was 
adhered to a solid surface and subjected to a simple shear flow.  The flow induced in the membrane
took the form of two vortices, rather than the 
simple overturning flow that would occur in a hemispherical droplet of one fluid in the background of 
an immiscible second fluid, without the membrane \cite{Dussan,Sugiyama}.  This difference is 
attributable to the incompressibility of the membrane, which restricts the flow field to
one that is two-dimensionally divergence free, i.e. area conserving, on the vesicle surface \cite{Woodhouse}.   

Since viscosity is the coefficient of proportionality 
between force per unit area on a surface and the adjacent shear rate, it is natural to ask whether
the experimental setup of V{\'e}zy, {\it et al.} \cite{Vezy} suggests a means to study membrane
fluid mechanics in detail.  To this end, we
describe here a method that quantifies the 
flows set up by shear of adherent vesicles, and, through a recent
calculation \cite{Woodhouse}, provides a means of determining membrane viscosity.  
The method uses Particle Image Velocimetry (PIV) to measure the
three-dimensional flows inside and outside vesicles, and particle tracking to monitor
the shear-induced movement of phase-separated domains within the membrane, in a  
microfluidic environment.  

Figure \ref{fig:droplet_diagram} shows the experimental setup: a 
vesicle of radius $R$, typically in the range of $10-40$ $\mu$m, adheres to the surface of a
microfluidic chamber in the presence of a flow with shear rate $\dot\gamma$.  
The chamber, typically $2$ mm wide and 
$200$ $\mu$m deep, is made from polydimethylsiloxane
(PDMS) by soft lithography and sealed with a glass coverslip that has been 
treated to promote vesicle adhesion. 
Vesicles were produced by standard methods of electroformation \cite{Angelova1992} in $100$ mM 
sucrose with or without $0.5$ $\mu$m microspheres (Invitrogen).
We chose lipid compositions to obtain two substantially different membrane viscosities. 
One composition gives primarily liquid-ordered ($L_{o}$) vesicles with a 
small fraction of liquid-disordered ($L_{d}$) phase at room temperature ($\sim 23^\circ C$): 
$40$ mol\% cholesterol (Sigma-Aldrich, MO, USA), $55$\% DPPC (dipalmitoylphosphatidylcholine), 
and $5$\% DiPhyPC  (diphytanoylphosphatidylcholine). DPPC, DOPC and DiPhyPC were purchased from 
Avanti Polar Lipids (Alabama, USA) and used without further purification. Vesicles containing 
primarily $L_{d}$ phase with a small fraction of gel domains were made from $85$\% DOPC 
and $15$\% DPPC.   The $L_{d}$ phases were labeled with $0.5$\% 
TexasRed-DPPE (Invitrogen). Coverslips were cleaned aggressively in NaOH and soaked 
in a solution of $0.001$\% polylysine for $30$ minutes for use with $L_{d}$ phase vesicles, or in $0.0005$\% 
polyethylenimine for $5$ minutes for use with $L_{o}$ phase vesicles. 
Vesicles were gently osmotically deflated by diluting into $130$ mM glucose and $10$ mM
HEPES shortly before loading into the chamber.

\begin{figure}[t]
\includegraphics[width = 1.0\columnwidth]{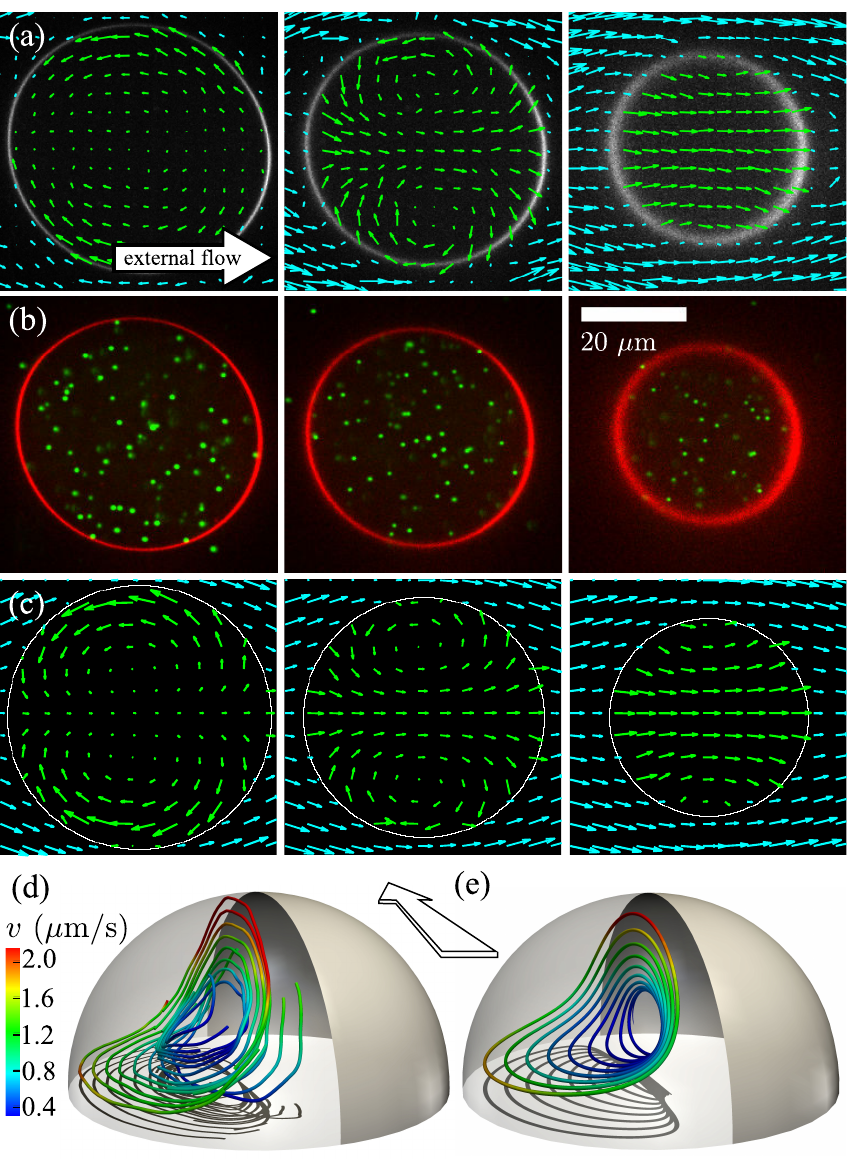}
\caption{(color online).   Flow fields inside an adhering vesicle in shear.  
(a)  Experimental 2D PIV velocity fields at heights $z/R=0.26, 0.47, 0.71$ above
coverslip.  
(b) Confocal slices at same fractional 
heights as (a) show  vesicle (red) containing fluorescent microspheres.   
(c) Theoretical 2D velocity fields \cite{Woodhouse} for a sheared hemispherical 
vesicle at $z/R=0.3, 0.5, 0.7$.   Interior and exterior PIV vectors in each panel of (a) and (c) have been 
rescaled for visual clarity.
(d) Experimental streamlines 
of the 3D velocity field obtained 
by integrating 2D flow fields, compared with theory (e).  Large arrows in (a), (d), and (e) indicate direction of
imposed shear flow.}
\label{fig:3D_flows}
\end{figure}

Measurements were made on a Zeiss Cell Observer spinning disk confocal microscope
with an electron-multiplied CCD camera (Evolve, Photometrics; $512\times 512$ pixels), using an 
NA 1.4/63X oil-immersion objective.   Flows were controlled by a syringe pump (PHD2000, Harvard Apparatus) and 
quantified by measuring far upstream from vesicles the speed of microspheres as a function of height above 
the coverslip.  Shear rates were typically in the range $1 \le \dot\gamma \le 6$ s$^{-1}$.
PIV was done with Matlab by adapting standard code \cite{piv_code}
to track small dilute tracers by finding 
the time-averaged velocity field \cite{Meinhart2000}. For 3D reconstruction, movies were recorded 
at $\sim\! 30$ frames per second at intervals of 2-3 $\mu$m throughout and above 
vesicles containing microspheres (Figure \ref{fig:3D_flows}b), giving 2D velocity field slices 
(Fig. \ref{fig:3D_flows}a).  From a stack of such slices a 3D 
velocity field was determined from the incompressibility relation. 
Figure \ref{fig:3D_flows}d shows a representative example of such streamlines. 
  
To understand the flows set up in and around the vesicle we distill the essential results of a 
recent calculation \cite{Woodhouse}.  Assume that the vesicle is a 
hemispherical cap of radius $R$ and origin $x=y=z=0$, adhered to the plane 
$z=0$, and let $(r,\theta,\phi)$ be spherical polars centered at the origin.
Given the fluid viscosity $\eta_-$ inside the vesicle ($r<R$), the membrane viscosity $\eta_m$, and the 
external fluid viscosity $\eta_+$ ($r>R$), we wish to find three velocity fields: $\mathbf{u}^-$ inside the vesicle,
the 2D flow $\mathbf{u}^m$ of the membrane, and $\mathbf{u}^+$ outside the vesicle 
(Fig. \ref{fig:droplet_diagram}a).  
The two external flows obey the unforced Stokes and incompressibility equations,
$\eta_\pm \nabla^2 \mathbf{u}^\pm - \boldsymbol{\nabla} p^\pm = \boldsymbol{0}$
and $\boldsymbol{\nabla} \boldsymbol{\cdot} \mathbf{u}^\pm = 0$, 
with far-field asymptotics $\boldsymbol{u}^+ \sim \dot\gamma z\hat{\boldsymbol{x}}$ 
as $r \rightarrow \infty$,  the no-slip condition $\boldsymbol{u}^\pm = \boldsymbol{0}$ on the plane
$\theta=\pi/2$, and no radial penetration,
$\boldsymbol{u}^\pm \boldsymbol{\cdot} \hat{\boldsymbol{r}} = 0$ at $r=R$.  The three velocities 
must be continuous across the membrane,
$\boldsymbol{u}^+ = \boldsymbol{u}^m = \boldsymbol{u}^- $ at $r=R$, and thus there
is the planar no-slip condition
condition $\boldsymbol{u}^m = \boldsymbol{0}$ at $\theta=\pi/2$.

\begin{figure}[t]
\includegraphics[width = 1.0\columnwidth]{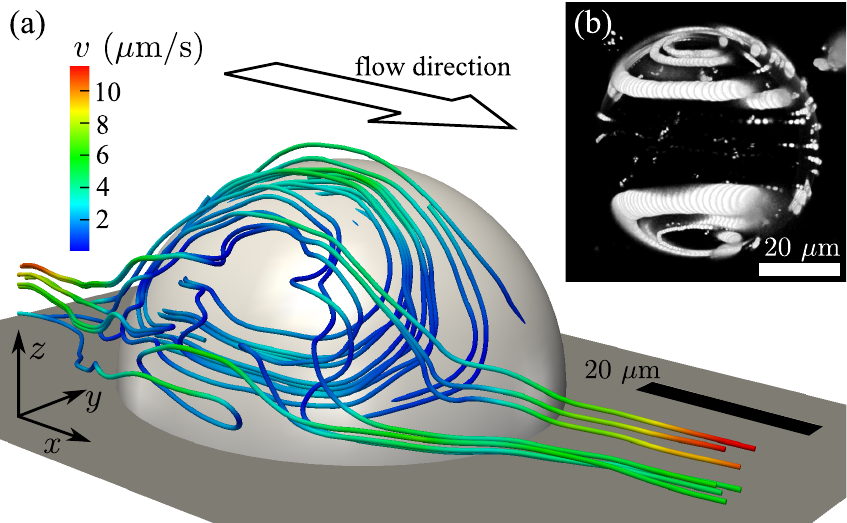}
\caption{(color online).   Membrane and external flows.  (a) Selected external streamlines along one side of an 
$L_{o}$ vesicle in shear flow, showing closed orbits above the surface.  (b) Time-lapse confocal stack of an 
$L_{o}$ vesicle, viewed from above, illustrating circulation of $L_{d}$ domains.}
\label{fig:vesicle_flows}
\end{figure}

Without a membrane, the bulk fluids' normal stresses would be continuous
at the interface, but a membrane can support tension, so the bulk stresses may be discontinuous.  
We assume that the membrane itself satisfies the Stokes equations and incompressibility,
$\hat{\boldsymbol{\nabla}} \boldsymbol{\cdot} 
\boldsymbol{u}^m = 0$, where $\hat{\boldsymbol{\nabla}}$ is the gradient operator constrained to the 
surface $r=R$.
This is a well-studied problem for Langmuir monolayers \cite{Schwartz,Lubensky1996}, but when the
membrane is curved there is a new contribution \cite{Henle} to the force balance relation
at the membrane involving its Gaussian curvature $K=R^{-2}$.  If
$\boldsymbol{e}^\pm_\parallel = e_{r \theta} \hat{\boldsymbol{\theta}} + e_{r \phi} 
\hat{\boldsymbol{\phi}}$ are the bulk fluids' in-plane normal rates-of-strain 
 the boundary condition is
\begin{equation}
\eta_m\left(\hat\nabla^2 \boldsymbol{u}^m + K\boldsymbol{u}^m\right)+
2\left[\eta_+\boldsymbol{e}^+_\parallel - 
\eta_- \boldsymbol{e}^-_\parallel\right]_{r=R} = \hat{\boldsymbol{\nabla}}\Pi.
 \label{eq:membrane_stressbc_noexpansion} 
\end{equation}

For the case $\eta_+=\eta_-$, the interior flows that emerge from this
calculation match closely those seen in experiment.  Cross-sectional profiles shown in 
Fig. \ref{fig:3D_flows}c at various elevations above the surface agree with the experimental profiles,
with significant counterflow both inside and {\it outside} the vesicle near its lateral edges, and over much of lowest
cross section.  The observed geometry of the internal streamlines (Fig. \ref{fig:3D_flows}d) 
follows that predicted theoretically (Fig. \ref{fig:3D_flows}e), and the maximum downstream
membrane speed is observed at the vesicle apex, as predicted.  The fluid motion external to the membrane and the
orbiting of domains within the membrane (Figs. \ref{fig:vesicle_flows}a,b) are both in agreement with theory. 
In addition to the circulating motion of the membrane domains we have observed over long periods of time their
gradual migration to the two vortex centers on either side of the vesicle midline, leaving a depleted region at the
apex (Fig. \ref{fig:vesicle_flows}b).  This appears to be an example
of the motion across streamlines described by Bretherton \cite{Bretherton}.
Note also the existence of closed streamlines outside the vesicle, as predicted \cite{Woodhouse}.

\begin{figure}
\includegraphics[width=1.0\columnwidth]{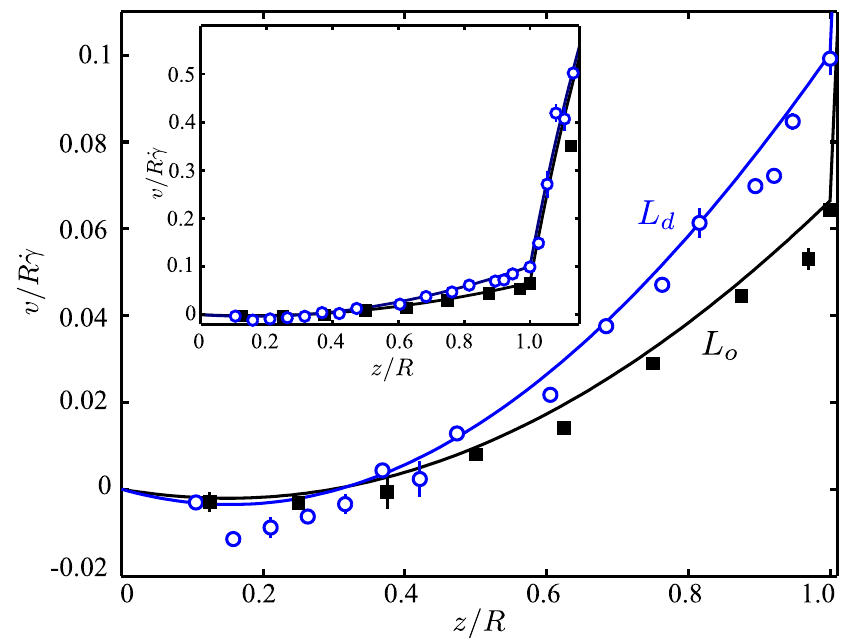}
\caption{(color online).  Downstream velocity profile through vesicle apex.  Data for 
an $L_{o}$ phase vesicle (black squares) and an $L_{d}$ phase vesicle (blue circles) are scaled by shear rate 
$\dot\gamma$ and vesicle radius $R$, displayed as a function of normalized height 
above the coverslip surface.
PIV experiments (symbols) show quantitative agreement with theoretical velocity predictions (lines) \cite{Woodhouse}. 
 Inset: extended 
plot to show the slope discontinuity at the membrane. }
\label{fig:velocity_profile}
\end{figure}

By plotting the downstream velocity as a function of $z$ through the vesicle apex (Fig. \ref{fig:velocity_profile}), a 
direct quantitative comparison can be made between theory and experiment.  
The discontinuity in the derivative of the fluid velocity at the membrane, set by
the gradient of the membrane tension through \eqref{eq:membrane_stressbc_noexpansion} 
is clearly seen in the downstream velocity as a function of $z$
through the vesicle apex (Fig. \ref{fig:velocity_profile}).  This provides perhaps the first direct measurement of tension 
gradients within bilayer membranes under shear.  For the vesicles composed primarily of $L_{o}$ phase 
the fluid velocity within the vesicle
is significantly lower than for $L_{d}$ vesicles as a direct consequence of the greater dissipation in the former,
as discussed further below.  Returning to the domain tracking in Figs. \ref{fig:droplet_diagram}c \& d, we observe 
smaller lateral thermal 
fluctuations in the $L_o$ vesicle due to its greater membrane viscosity.

An important empirical result of the calculation \cite{Woodhouse} is that the speed 
$v_{0} = \vert\boldsymbol u^m\vert$ of the membrane at the
apex of the vesicle has a simple dependence on $r_\pm \equiv \eta_m/R \eta_\pm$,
the non-dimensional form of the `Saffman-Delbr{\"u}ck' lengths 
$\ell_\pm \equiv \eta_m/\eta_\pm$ \citep{SaffmanDelbruck, Henle}:
$R\dot\gamma/v_{0} = Ar_{+}/r_{-} + Br_{+} + C~$,
where $A,B,C$ are known constants.
When the inner and outer viscosities are equal (the sucrose and glucose solutions used 
have viscosities within 4\% of each other) this result further reduces to the
simple linear dependence,
\begin{equation}
\frac{R\dot\gamma}{v_{0}} \simeq 7.86+ 4.72 \frac{\eta_{m}}{\eta_{w} R},
\label{tada}
\end{equation}
where $\eta_w$ is the viscosity of water. Since $R\dot\gamma$ is the fluid velocity in the shear profile
in the absence of the vesicle, we term the quantity $R\dot\gamma/v_0$ the {\it velocity attenuation ratio}.  
Its predicted variation with the Saffman-Delbr{\"u}ck length provides a means of determining
$\eta_m$ from the apex velocity for a range of vesicle radii.  

\begin{figure}[t]
\includegraphics[width = 1.0\columnwidth]{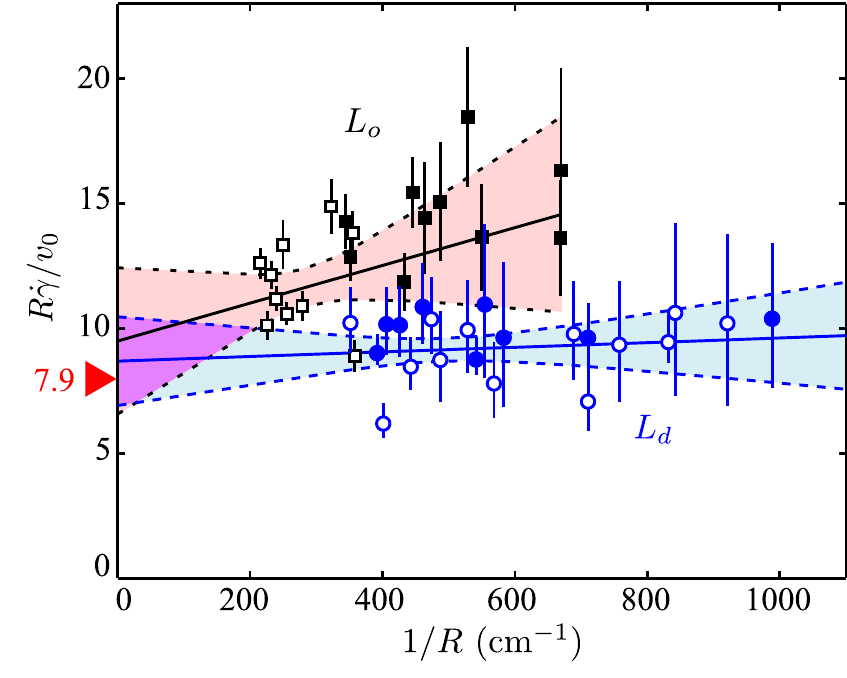}
\caption{(color online).  Test of the predicted velocity attenuation relation.  Scaled membrane velocity $v_0$ at
vesicle apex for $L_{o}$ phase (black squares) and $L_{d}$ phase (blue circles) as a function of inverse vesicle 
radius, for $\sim\! 20$ vesicles in each phase. For each data point, a given vesicle was observed
at  2-3 different shear rates to confirm linearity of $v_0$ with $\dot\gamma$, and the results averaged.  
Closed symbols indicate vesicles whose height-to-radius ratio was within $10\%$ of unity.  Open symbols deviate by no 
more than $60\%$.  Black and blue lines indicate weighted least squares fits to the linear relation (\ref{tada}), 
the slopes of which yield the associated membrane viscosities.
Pink and blue shaded areas indicate $95\%$ confidence intervals, purple indicates overlap.  Red arrow indicates 
predicted intercept. }
\label{fig:axisymmetric_amplitudes}
\end{figure}

This analysis shows that in the limit of large vesicle radius or vanishing membrane viscosity the
velocity attenuation ratio is the constant $\sim 7.9$.  (A viscosity contrast changes the value of the intercept, but the
relationship persists in character.)  This is a significant and purely geometric effect: the
constraint of lateral incompressibility on the membrane velocity field drastically reduces the fluid flow, even in the 
absence of any internal membrane dissipation.  Values of membrane viscosities reported previously 
\cite{Dimova,Petrov} have ranged
from $2-6$ $\mu$Poise$\cdot$cm in disordered lipid phases and from studying dynamic fluctuations near a 
compositional critical point in membranes \cite{Camley,HonerkampSmith}.  
Values at the lower end would change the
velocity attenuation ratio by $\sim 0.5$ for the smallest vesicles that can be studied by this method 
($\sim 10$ $\mu$m), an increment that is below the noise limit
for our measurements.  This is seen in the data shown in blue in Fig. \ref{fig:axisymmetric_amplitudes} for 
$L_{d}$ membranes.  A weighted least squares fit yields an intercept of $8.69\pm 2.73$, in excellent agreement
with the predicted value, and a negligible membrane 
viscosity ($\eta_m=1.9\pm 11$ $\mu$Poise$\cdot$cm).   This lipid composition thus serves mainly as a control to test
the zeroth order velocity attenuation due to membrane incompressibility.  On the other hand, the data for
the $L_{o}$ phase vesicles shows a large change in the attenuation ratio and yields a membrane viscosity of
$\eta_m=15.7\pm 9.9$ $\mu$Poise$\cdot$cm (with a consistent intercept of $9.50\pm 1.41$).   
The overlapping $95\%$ confidence intervals for extrapolations of both data sets to $R\to \infty$ indicate
clear statistical consistency between the two and with the theoretical prediction.

Finally, we note that recent theoretical work shows that the inclusion of gel domains into a fluid membrane would 
increase its viscosity to a degree that depends on several factors, including the
ratio of domain radius $a$ to the Saffman-Delbr{\"u}ck lengths $\ell_\pm$,
and the area fraction of domains $\phi$ \cite{Henle2009}. Domain area fractions for
both compositions used here
were low, $1-2\%$ for $L_{o}$ phase vesicles and $1-10\%$ for $L_{d}$
phase vesicles, and the dilute solution analysis in \cite{Henle2009} suggests that
our measurements may overestimate the true viscosity by about $4\%$ in
the $L_{o}$ phase and $14\%$ in the $L_{d}$ phase.
Average domain separations were $\sim 2-10$ $\mu$m
for both compositions, so our $L_{d}$ membranes
did not fully meet the dilute condition required by theory,
and hydrodynamic interactions between domains might
increase further the apparent viscosity. However, we saw no systematic trend in measured
viscosity of individual vesicles with $\phi$ or domain separation over the small
range studied.

In summary, we have presented the first detailed fluid mechanical measurements of the
flows inside, on, and around lipid bilayer vesicles under controlled conditions of fluid shear.  Detailed 
analysis of those flow fields for low-viscosity membranes confirms quantitatively a theoretically predicted geometric
velocity attenuation effect, and it reveals the scale of membrane viscosity necessary to significantly affect shear-driven 
flows. The combination of techniques 
described here may prove useful in the study of more complex systems involving membranes under shear, 
such as those found in large eukaryotic cells and perhaps in contexts within developmental biology, where 
the membranes may be more tightly coupled to cytoskeletal filaments and the internal cellular rheology may be
non-Newtonian.

We thank S. Ganguly, S.L. Keller, and M. Polin for contributions at an early stage of this work and many
valuable discussions.  This work was supported by the Leverhulme Trust, the Engineering and Physical 
Sciences Research Council, and the European Research Council Advanced Investigator Grant 247333 (R.E.G.).


\begin{thebibliography}{99}

\bibitem{SaffmanDelbruck} P.G. Saffman and M. Delbr{\"u}ck, Proc. Natl. Acad. Sci. USA
{\bf 72},  3111 (1972); P.G. Saffman, J. Fluid Mech. {\bf 73}, 593 (1976).

\bibitem{HPW} B.D. Hughes, B.A. Pailthorpe, and L.R. White, J. Fluid Mech. 
{\bf 110}, 349 (1981).

\bibitem{KellerSkalak} S.R. Keller and R. Skalak, J. Fluid Mech. {\bf 120}, 27 (1982).

\bibitem{NoguchiGompper} H. Noguchi and G. Gompper, Phys. Rev. Lett. {\bf 93},
258102 (2004); {\bf 98}, 128103 (2007).

\bibitem{Kantsler} V. Kantsler and V. Steinberg, Phys. Rev. Lett. {\bf 95}, 258101 (2005); {\bf 96}, 036001 (2006).

\bibitem{Seifert1993} U. Seifert, S.A. Langer, Europhys. Lett. {\bf 23}, 71 (1993). 

\bibitem{Dimova} R. Dimova, C. Dietrich, A. Hadjiiski, K. Danov, and B. Pouligny, 
Eur. Phys. J. B {\bf 12}, 589 (1999); K.D. Danov, R. Dimova, and B. Pouligny,
Phys. Fluids {\bf 12}, 2711 (2000).

\bibitem{Petrov} E.P. Petrov, R. Petrosyan, and P. Schwille, Soft Matter {\bf 8}, 
7552 (2012).

\bibitem{Cicuta2007} P. Cicuta, S.L. Keller, and S.L. Veatch, J. Phys. Chem. B {\bf 111}, 3328 (2007).

\bibitem{Camley} B.A. Camley, C. Esposito, T. Baumgart, and F.L.H. Brown,
Biophys. J. {\bf 99}, L44 (2010).

\bibitem{HonerkampSmith} A. R. Honerkamp-Smith, B.B. Machta, and S.L. Keller, Phys. Rev. Lett. 
{\bf 108}, 265702 (2012).

\bibitem{Fuller2012} G.G. Fuller and J. Vermant, Annu. Rev. Chem. Biomol. Eng. {\bf 3}, 519 (2012).

\bibitem{Levine}  R. Shlomovitz, A.A. Evans, T. Boatwright, M. Dennin, and A.J. Levine, Phys. Rev. Lett. 
{\bf 110}, 137802 (2013).

\bibitem{Sickert2007} M. Sickert, F. Rondelez, and H.A. Stone, Europhys. Lett. {\bf 79}, 66005 (2007).

\bibitem{Sadoughi2013} A.H. Sadoughi, J.M. Lopez, and A.H. Hirsa, Phys. Fluids {\bf 25}, 032107 (2013).

\bibitem{Protoplasma} J. Verchot-Lubicz and R.E. Goldstein, Protoplasma {\bf 240}, 99 (2009).

\bibitem{Shimmen04} T. Shimmen and E. Yokota, Curr. Opin. Cell Biol. {\bf 16}, 
68 (2004).

\bibitem{Goldstein2008} R.E. Goldstein, I. Tuval and J.W. van de Meent, Proc. Natl. Acad. 
Sci. USA {\bf 105}, 3663 (2008).

\bibitem{chara_jfm} J.-W. van de Meent, A.J. Sederman, L.F. Gladden, and R.E. Goldstein,
J. Fluid Mech. {\bf 642}, 5 (2010).

\bibitem{Wolff} K. Wolff, D. Marenduzzo, and M.E. Cates, J. Roy. Soc. Interface {\bf 7}, 1398 (2012).

\bibitem{Vezy} C. V{\'e}zy, G. Massiera, and A. Viallat, Soft Matter {\bf 3}, 844 (2007).

\bibitem{Lorz} B. Lorz, R. Simson, J. Nardi, and E. Sackmann, Europhys. Lett. 
{\bf 51}, 468 (2000).

\bibitem{Dussan} E.B. Dussan V., J. Fluid Mech. {\bf 174}, 381 (1987).

\bibitem{Sugiyama} K. Sugiyama and M. Sbragaglia, J. Eng. Math. {\bf 62}, 35 (2008).

\bibitem{Woodhouse} F.G. Woodhouse and R.E. Goldstein, J. Fluid Mech. {\bf 705}, 165 (2012).

\bibitem{Angelova1992} M.I. Angelova, S. Soleau, P. Meleard, J.F. Faucon, and P. Bothorel, 
Prog. Colloid Polym. Sci. {\bf 89}, 127 (1992).

\bibitem{piv_code} Available at http://www.oceanwave.jp/softwares/mpiv.

\bibitem{Meinhart2000} C.D. Meinhart, S.T. Wereley, and J.G. Santiago, J. Fluid Eng. {\bf 122}, 285 (2000).

\bibitem{Schwartz} D.K. Schwartz, C.M. Knobler, and R. Bruinsma, Phys. Rev. Lett. {\bf 73}, 2841 (1994).

\bibitem{Lubensky1996} D.K. Lubensky and R.E. Goldstein, Phys. Fluids {\bf 8} 4, 843 (1996). 

\bibitem{Henle} M.L. Henle, R. McGorty, A.B. Schofield, A.D. Dinsmore, and A.J. Levine,
Europhys. Lett. {\bf 84}, 48001 (2008); M.L. Henle and A.J. Levine, Phys. Rev. E
{\bf 81}, 011905 (2010).

\bibitem{Bretherton} F.P. Bretherton, J. Fluid Mech. {\bf 14}, 284 (1962);  See also G.B. Jeffery, 
Proc. Roy. Soc. London A {\bf 102}, 161 (1922).

\bibitem{Henle2009} M.L. Henle and A.J. Levine, Phys. Fluids {\bf 21}, 033106 (2009). 

\end{thebibliography}
\end{document}